\documentclass[aps,prd,twocolumn,groupedaddress,asmsymb,amsmath,amssymb,longbibliography]{revtex4-1}

\usepackage{graphicx}
\usepackage{calligra}
\usepackage{upgreek}
\usepackage{color}
\usepackage{caption}
\usepackage{stackrel}
\usepackage[usenames,dvipsnames]{xcolor}

\numberwithin{equation}{section}

\DeclareMathAlphabet{\mathpzc}{OT1}{pzc}{m}{it}

\newcommand{\be}{\begin{equation}}
\newcommand{\ee}{\end{equation}}
\newcommand{\bea}{\begin{eqnarray}}
\newcommand{\eea}{\end{eqnarray}}
\newcommand{\lb}{\label}

\newcommand{\bu}{{\bf u}}

\newcommand{\bx}{{\bf x}}

\newcommand{\br}{{\bf r}}

\newcommand{\btau}{{\mbox{\boldmath $\tau$}}}

\newcommand{\grad}{{\mbox{\boldmath $\nabla$}}}
\newcommand{\bdot}{{\mbox{\boldmath $\cdot$}}}
\newcommand{\bdots}{{\mbox{\boldmath $:$}}}
\newcommand{\bzed}{{\mbox{\boldmath $0$}}}

\begin{document}

%\preprint{}

\title{Review of the Onsager ``Ideal Turbulence'' Theory}
%and Dissipative Anomalies}


\author{Gregory L. Eyink${\,\!}^{1,2}$}
%\thanks{}
\affiliation{${\,\!}^1$Department of Applied Mathematics \& Statistics, The Johns Hopkins University, Baltimore, MD, USA, 21218}
\affiliation{${\,\!}^2$Department of Physics \& Astronomy, The Johns Hopkins University, Baltimore, MD, USA, 21218}


\date{\today}

\begin{abstract}
In his famous undergraduate physics lectures, Richard Feynman remarked 
about the problem of fluid turbulence: ``Nobody in physics has really been able to analyze it 
mathematically satisfactorily in spite of its importance to the sister sciences'' \cite{feynman1965feynman}. 
This statement was already false when Feynman made it. Unbeknownst to him, Lars Onsager decades 
earlier had made an exact mathematical analysis of the high Reynolds-number limit of incompressible 
fluid turbulence, using a method that would now be described as a non-perturbative renormalization 
group analysis and discovering the first ``conservation-law anomaly'' in theoretical physics. 
Onsager's results were only cryptically announced in 1949 and he never published any of his 
detailed calculations. Onsager's analysis was finally rescued from oblivion and reproduced 
by this author in 1992. The ideas have subsequently been intensively developed in the mathematical 
PDE community,  where deep connections emerged with John Nash's work on isometric embeddings. 
Furthermore, Onsager's method has more recently been successfully applied to new physics problems, 
such as compressible fluid turbulence and relativistic fluid turbulence, yielding many novel testable predictions.  
This note will explain Onsager's exact analysis of incompressible turbulence using modern ideas 
on renormalization group and conservation-law anomalies, and it will also very briefly review subsequent 
developments.
\end{abstract}

%\pacs{?????}
%\keywords{}

\maketitle

%\section{}
%\section{}

\section{Introduction}\label{sec:II} 

Onsager's several contributions to the theory of turbulence have already been reviewed from 
a history of science  point of view \cite{eyink2006onsager}. This note is instead intended 
to give a busy, working physicist a concise, accurate and painless explanation of Onsager's  theory 
of ``ideal turbulence'' for a low Mach-number fluid, described by the incompressible 
Navier-Stokes equation
\be \partial_t\bu+(\bu\bdot\grad_\bx)\bu=-\grad_\bx p+\nu\triangle\bu, 
\quad \grad_\bx\bdot\bu=0. \lb{II1} \ee
For previous physics explanations of the Onsager theory, see 
\cite{eyink1995local} and, for an extended, pedagogical presentation, the course notes 
of \cite{eyink2010notes}. None of these works, however, explain the theory systematically 
from the point of view of renormalization group, as we do here.  We shall avoid full 
rigorous details (which can be readily found in the cited papers) and instead focus 
on the intuitive ideas, exact calculations, and essential estimates. 

\section{Divergences and Regularization}\lb{secIIa}

The key empirical fact underlying the Onsager theory is the non-vanishing of turbulent 
energy dissipation in the zero-viscosity limit. This was first suggested on a 
semi-phenomenological basis by Taylor \cite{taylor1917observations}, 
who argued that energy could be dissipated 
``in fluid of infinitesimal viscosity''.  More properly, the phenomenon occurs in the limit of high Reynolds 
numbers. When the equations are rescaled by characteristic large length $L$ and velocity $U,$ 
then in terms of the dimensionless variables
\be \hat{\bx}=\bx/L, \quad \hat{t}=t/(L/U), \quad \hat{\bu}=\bu/U,
\quad \hat{p}=p/U^2, \lb{II2} \ee
the Navier-Stokes equation assume the similarity form 
\be \hat{\partial}_{t}\hat{\bu}+(\hat{\bu}\bdot\hat{\grad})\hat{\bu}=
-\hat{\grad} \hat{p}+ \frac{1}{Re}\hat{\triangle}\hat{\bu}, 
\quad \hat{\grad}\bdot\hat{\bu}=0 \lb{II3} \ee
with $Re=UL/\nu=1/\hat{\nu}$ the Reynolds number. Hereafter we omit the hats $\widehat{(\cdot)}$
and understand that the limit $\nu\to 0$ is really to be interpreted as the limit $Re\to \infty.$
Laboratory experiments \cite{sreenivasan1984scaling,pearson2002measurements}
and numerical simulations 
\cite{sreenivasan1998update,kaneda2003energy}
both confirm that the kinetic energy dissipation rate 
\be \varepsilon(\bx,t):=\nu|\grad_\bx\bu(\bx,t)|^2  \lb{II4} \ee
has space-average converging in the limit as $\nu\to 0$ and not vanishing: 
$\langle\varepsilon(t)\rangle\to \langle\varepsilon_\star(t)\rangle>0.$
It is furthermore observed in experiment that 
when integrated over small balls or cubes in space the high-$Re$ limit $\varepsilon_\star(\bx,t)$ 
defines a positive measure with multifractal scaling  
\cite{meneveau1987multifractal,meneveau1991multifractal}. 

The most obvious requirement for such a non-vanishing limit of dissipation is that space-gradients of velocity 
must diverge, $\grad_\bx\bu\to\infty,$ as $\nu\to 0.$ This is a short-distance/ultraviolet (UV) divergence 
in the language of quantum field-theory, or what Onsager himself termed a ``violet catastrophe'' 
\cite{onsager1945distribution}. 
The inviscid limit for turbulent fluids is analogous to a ``continuum'' or ``critical'' limit in quantum field-theory, 
where a scale-invariant regime is expected and similar UV divergences are encountered. 
Since the fluid equations of motion (\ref{II1}) contain diverging gradients, they become ill-defined in the limit. 
In order to develop a dynamical description which can be valid even as $\nu\to 0,$ some regularization 
of this divergence must be introduced. Here we shall employ a simple coarse-graining or 
``block-spin'' regularization, defining a coarse-grained velocity field at length-scale $\ell$ by local
space-averaging: 
\be \overline{\bu}_\ell(\bx,t)=\int d^3x\  G_\ell(\br)\, \bu(\bx+\br,t).   \lb{II5} \ee 
The coarse-graining kernel $G_\ell(\br)=\ell^{-3}G(\br/\ell)$ may be chosen rather freely, subject 
to general constraints that the function $G$ be
\bea 
&& \qquad G(\br)\geq 0 \quad \mbox{ (non-negative) }\cr
&& \int d^3r\ G(\br)=1 \quad \mbox{ (normalized) } \cr
&& \int d^3r\ \br\, G(\br)=\bzed  \quad \mbox{ (centered) } \cr
&& \int d^3r\ |\br|^2\, G(\br)=1 \quad \mbox{ (unit variance) } 
\lb{II6} \eea 
Also, we require that $G$ be both smooth and rapidly decaying in space, e.g. $G\in C^\infty_c({\mathbb R}^3),$
the space of infinitely-differentiable functions with compact support. We also for convenience assume isotropy, or 
$G=G(r)$ with $r=|\br|,$ so that $\int d^3r\ r_i r_j \, G(\br)=(1/3)\delta_{ij}.$ We note in passing that an 
identical coarse-graining operation is employed in the  ``filtering approach'' to turbulence advocated 
by Germano \cite{germano1992turbulence}, but with a different motivation than regularization of divergences.  
For us, coarse-graining is a convenient choice as a regularizer for many purposes, but not the only one 
possible and not always even adequate (e.g. see section VII-A of \cite{eyink2018kinetic}). 

The coarse-graining operation \eqref{II5} clearly regularizes gradients, so that 
$\grad_\bx\overline{\bu}_\ell$ remains finite as $\nu\to 0$ for any fixed length $\ell>0.$ This may be shown formally 
using the simple identity
\be \grad_\bx\overline{\bu}_\ell(\bx,t)=-\frac{1}{\ell}\int d^3r\  (\grad G)_\ell(\br)\, \bu(\bx+\br,t),  
\lb{II7} \ee 
which by Cauchy-Schwartz inequality yields the bound $|\grad_\bx\overline{\bu}_\ell(\bx,t)|\leq 
(1/\ell)\sqrt{C_\ell\int d^3r \, |\bu(\br,t)|^2}$ with constant $C_\ell=\int d^3r\ |(\grad G)_\ell(\br)|^2.$
Thus, the coarse-grained gradient is bounded as long as the total energy remains finite as $\nu\to 0$
(which, for instance, is necessarily true for freely-decaying turbulence with no stirring). The price 
of this regularization is that a new, arbitrary length-scale $\ell$ has been introduced.  Although the 
{\it description} of turbulent phenomena will depend very strongly on a particular choice, it is clear 
that no objective physical fact can depend upon the arbitrary scale $\ell.$  The coarse-graining performed in  \eqref{II5} is 
a purely passive operation, which corresponds to observing the fluid at spatial resolution $\ell,$ 
whereas objective facts %such as the persistent decay of kinetic energy in the limit $\nu\to 0$ 
cannot depend 
upon the ``eyesight'' of the observer. This situation exactly parallels that in quantum field-theory, where regularization 
of ultraviolet divergences in weak-coupling perturbation theory introduces a new arbitrary momentum scale 
$\mu$ at which coefficients of the renormalized theory are defined and upon which objective physics 
cannot depend.  This is the statement of what is called ``renormalization-group invariance'' in quantum 
field-theory and condensed matter physics 
and renormalization group (RG) methods can be understood as the systematic exploitation 
of the invariance of the physics to changes of this arbitrary regularization scale 
(see \cite{gross1976applications}, section 4). 

We follow a similar strategy here. An energy dissipation rate which is non-vanishing in the 
limit as $\nu\to 0$ implies that energy decrease over a fixed interval of time $[0,t]$ will persist 
in the inviscid limit. On the other hand, an observer at the coarse-grained scale $\ell$ can only 
miss some kinetic energy of smaller eddies, since by convexity 
\be \frac{1}{2}|\overline{\bu}_\ell(\bx,t)|^2\leq \frac{1}{2}\overline{\left(\left|\bu(\bx,t)\right|^2\right)}_\ell, 
\lb{II8} \ee 
and it then follows that %the ``myopic observer''  can only miss some kinetic energy, that is, 
$E_\ell(t):=(1/2)\int d^3x\, |\overline{\bu}_\ell(\bx,t)|^2\leq (1/2)\int d^3x\, |\bu(\bx,t)|^2=E(t).$ If kinetic 
energy continues to decay even in the limit as $\nu\to 0,$ then such persistent energy decay 
must also be seen by the ``myopic'' observer who observes fluid features only at space-resolution 
$\ell.$  As we now show, however, the persistent energy decay observed at the fixed length-scale 
$\ell$ with $\nu\to 0$ is not due to molecular viscosity acting directly at those scales. 

\section{Coarse-Grained Euler and Energy Cascade}\lb{sec:IIb} 

The equation obeyed by the coarse-grained velocity field defined in \eqref{II5} is easily found to be
\be \partial_t\overline{\bu}_\ell+\grad_\bx\bdot\overline{(\bu\bu)_\ell}=
-\grad_\bx\overline{p}_\ell+\nu\triangle\overline{\bu}_\ell, \quad \grad_\bx\bdot\overline{\bu}_\ell=0,
\lb{II9} \ee 
because the coarse-graining operation commutes with all space- and time-derivatives. By writing
\be \nu\triangle\overline{\bu}_\ell(\bx,t)=-\frac{\nu}{\ell}\int d^3r\  (\grad G)_\ell(\br)\bdot\grad_\bx\bu(\bx+\br,t),  
\lb{II10} \ee 
and by again applying Cauchy-Schwartz inequality, one obtains \footnote{A much sharper estimate is obtained 
from the identity
$\nu\triangle\overline{\bu}_\ell= (\nu/\ell^2)\int d^3r\  (\triangle G)_\ell(\br)\delta \bu(\br),$ 
which shows that $\nu\triangle\overline{\bu}_\ell=O(\nu \delta u(\ell)/\ell^2),$ vanishing as 
$\nu\to 0$ for fixed $\ell$ or as $\ell\to\infty$ for fixed $\nu.$ The viscous diffusion term is 
thus an ``irrelevant variable'' in the technical RG sense. We presented the weaker estimate 
\eqref{II11} in the main text because it illustrates the strategy that we shall employ to 
estimate the direct effect of collisions in the ``inertial range'' of kinetic plasma turbulence.} 
\be \left|\nu\triangle\overline{\bu}_\ell(\bx,t)\right|\leq 
\sqrt{\frac{\nu}{\ell} C_\ell \int d^3r\  |(\grad G)_\ell(\br)|^2\varepsilon(\bx+\br,t)}. \lb{II11} \ee 
Since the integral $\int d^3r\  |(\grad G)_\ell(\br)|^2\varepsilon(\bx+\br,t)$ as $\nu\to 0$\\ converges  
to  $\int d^3r\  |(\grad G)_\ell(\br)|^2\varepsilon_\star(\bx+\br,t),$ then the viscous diffusion 
term in \eqref{II9} has an upper bound $O(\sqrt{\nu})$ and thus vanishes for fixed $\ell$ in the 
limit $\nu\to 0.$ The result is a simplified set of dynamical equations 
\be \partial_t\overline{\bu}_\ell+\grad_\bx\bdot\overline{(\bu\bu)_\ell}=
-\grad_\bx\overline{p}_\ell, \quad \grad_\bx\bdot\overline{\bu}_\ell=0,
\lb{II12} \ee 
at the range of scales $\ell$ where the viscosity term is negligible. This set of length-scales $\ell$
constitute what is called the ``inertial-range'' in the turbulence literature, where the direct action 
of viscosity is vanishingly small. The dynamical equations \eqref{II12} are what we shall term 
``coarse-grained Euler equations'' at the length-scale $\ell$. This notion formalizes 
the intuitive idea that the inertial-range eddies should have their dynamics governed 
by the ideal Euler equations \footnote{For a traditional statement of this idea, consider for example 
the textbook of Landau \& Lifschitz \cite{landau1959fluid}, \S 31:  
``We therefore conclude that, for the large eddies which 
are the basis of any turbulent flow, the viscosity is unimportant and may be equated to zero, so that 
the motion of these eddies obeys Euler's equation. In particular, it follows from this that there is no 
appreciable dissipation of energy in the large eddies''. The latter statement is only true, however, 
for the direct viscous dissipation of kinetic energy at inertial-range scales, whereas the energy 
in eddies at those scales, in fact, must be dissipated. The rate of decrease of energy for free-decay
or rate of power-input for forced turbulence are objective facts that cannot depend upon the 
resolution of eddies in the inertial-range}. If a strong limit $\bu_\star=\lim_{\nu\to 0} \bu$
exists, then the coarse-grained equations \eqref{II12} hold for $\bu_\star$ at any length-scale $\ell>0$
and the inviscid limit field is what in mathematics  is called a ``distributional'' or ``weak'' 
solution of the incompressible Euler equations. See Propositions 1 and 2 in \cite{drivas2018onsager}. 

Although Euler equations hold in the coarse-grained sense within the inertial-range of scales,
the energy contained within those eddies is not conserved in time. As the argument below
\eqref{II8} demonstrates, if energy dissipation persists in the limit as $\nu\to  0,$ then energy 
must also decay for the observer with space-resolution $\ell,$ even though for sufficiently small 
$\nu$ the fluid motions at the fixed scale $\ell$ are described by ``Euler equations". The resolution 
of this seeming paradox is that the statement that the velocity field $\bu$ satisfies 
the ``coarse-grained Euler equations'' \eqref{II12} at scale $\ell$ is quite different 
from the statement that the coarse-grained velocity $\overline{\bu}_\ell$ satisfies the 
Euler equations in the naive sense. To make this point very clearly, 
we introduce the ``turbulent'' or ``sub-scale'' stress-tensor
\be \btau_\ell(\bu,\bu) = \overline{(\bu\bu)_\ell} -\overline{\bu}_\ell\overline{\bu}_\ell \lb{II13} \ee
in terms of which the ``coarse-grained Euler equations'' \eqref{II12} may be equivalently written as 
\cite{eyink1995local} 
\be \partial_t\overline{\bu}_\ell+\grad_\bx\bdot\left(\overline{\bu}_\ell\overline{\bu}_\ell+\btau_\ell\right)=
-\grad_\bx\overline{p}_\ell+\nu\triangle\overline{\bu}_\ell, \quad \grad_\bx\bdot\overline{\bu}_\ell=0,
\lb{II14} \ee 
Only for $\btau_\ell\equiv \bzed$ does \eqref{II14} correspond to $\overline{\bu}_\ell$ satisfying the 
incompressible Euler equations in the naive sense. Once one understands that the inertial-range 
eddies satisfy the ``Euler equations'' only in the coarse-grained sense of \eqref{II12} or \eqref{II14},
then it is no mystery how the energy is dissipated at those scales. The local kinetic energy balance
at length-scales $\ell$ in the inertial-range is easily calculated to be 
\bea 
&& \partial_t\left(\frac{1}{2}|\overline{\bu}_\ell|^2\right)
+\grad_\bx\bdot\left[\left( \frac{1}{2}|\overline{\bu}_\ell|^2+\overline{p}_\ell\right)\overline{\bu}_\ell
+\btau_\ell\bdot\overline{\bu}_\ell\right] 
=-\Pi_\ell\cr
&& \lb{II15} \eea
where the quantity on the right side of the equation, 
\be \Pi_\ell(\bx,t) = -\grad_\bx\overline{\bu}_\ell(\bx,t)\,\bdots\,\btau_\ell(\bx,t),  \lb{II16} \ee 
is the ``deformation work''  \cite{tennekes1972first} of the large-scale strain acting 
against small-scale stress, or the ``energy flux'' from resolved scales $>\ell$ to 
unresolved scales $<\ell.$ The mechanism of loss of energy by  the inertial-range eddies is thus ``energy cascade'', a 
term first used in this connection by Onsager \cite{onsager1945distribution}. 

Note that the energy flux defined
in \eqref{II16} is a spatially local version of the standard concept of ``spectral energy flux'' $\Pi(k,t)$ (e.g. see Frisch (1995),
section 6.2.2). Indeed, it is not difficult to show \footnote{Note that  Parseval's equality and definition
of energy spectrum $E(k,t)$ give $(1/|V|)\int_V d^3x\ (1/2)|\bar{\bu}_\ell(\bx,t)|^2 = \int_0^\infty dk\ |\hat{G}(k\ell)|^2\, E(k,t)$
and likewise the definition of the power-input spectrum $F(k,t)$ gives 
$(1/|V|)\int_V d^3x\ \bar{\bu}_\ell(\bx,t)\bdot \overline{{\bf f}}_\ell(\bx,t)= \int_0^\infty dk\ |\hat{G}(k\ell)|^2\, F(k,t).$
The standard definition of the spectral flux $\Pi(k,t)$
gives $\partial_t E(k,t)=-\partial_k \Pi(k,t)+F(k,t),$ which easily yields the stated result \eqref{II17}}
that the two fluxes are related by 
\be \frac{1}{|V|}\int_V d^3x\ \Pi_\ell(\bx,t) = \int_0^\infty dk\ P_\ell(k) \, \Pi(k,t) \lb{II17} \ee
where, for any isotropic kernel $G(r)$ with 3D Fourier transform $\widehat{G}(k),$
the formula 
\be  P_\ell(k) = -\frac{d}{dk} |\widehat{G}(k\ell)|^2 \lb{II18} \ee 
defines a distribution function satisfying $\int_0^\infty dk\, P_\ell(k)=1$
and $P_\ell(k)\geq 0$ for standard kernels with $|\widehat{G}(k)|^2$ decaying 
monotonically in wave-number. Intuitively, the flux $\Pi_\ell(\bx,t)$ is well-localized 
in physical space and $\Pi(k,t)$ is well-localized in Fourier space, but their respective 
averages over space and wavenumber agree.  Note that the width of distribution 
$P_\ell(k)$ in $k$-space is $\Delta k\sim 1/\ell$, consistent with the uncertainty principle of 
Fourier analysis, $\Delta k\,\Delta x\sim 1$. 
When time-average spectral flux is constant in a long range of wavenumbers $k$
where $\langle \Pi(k)\rangle=\varepsilon,$  then identity 
(\ref{II17}) implies $\langle \Pi_\ell\rangle=\varepsilon$ for $\ell\sim 1/k.$ 

\section{Velocity-Increments and Singularities}\lb{sec:IIc} 

The turbulent stress tensor $\btau_\ell(\bu,\bu)$ defined in \eqref{II13} is not a simple functional 
of the resolved velocity $\overline{\bu}_\ell.$ This can be seen by recasting the 
dynamical equation \eqref{II1} as a path-integral over an ensemble of velocities $\bu$, 
by assuming either random initial data $\bu_0$ or by adding to the righthand side of the 
momentum balance a random stirring force ${\bf f}.$ Writing $\bu=\overline{\bu}_\ell+\bu_\ell'$ 
and integrating out the small-scale field $\bu_\ell'$ yields a reduced path-integral 
for $\overline{\bu}_\ell$. This new path-integral corresponds exactly to the coarse-grained
equation \eqref{II14}, where the stress $\btau_\ell$ produced by integrating out $\bu_\ell'$ is 
a highly complicated functional of $\overline{\bu}_\ell,$ with transcendental nonlinearity, 
long-term memory, and intrinsic stochasticity (e.g. see \cite{eyink1996turbulence}). 
This is not surprising,
since Wilson-Kadanoff  RG procedures typically lead to highly complicated effective actions 
in the path-integrals for ``block-spin'' fields $\overline{\bu}_\ell.$ This lack of a simple 
expression for $\btau_\ell(\bu,\bu)$ in terms of the resolved velocity $\overline{\bu}_\ell$
is what is termed the ``closure problem'' of turbulence theory. For engineering modelling
by the ``Large-Eddy Simulation'' (LES) method, the primary problem is to develop
suitable model expressions $\btau_\ell^M[\overline{\bu}_\ell]$ that are closed in terms of 
$\overline{\bu}_\ell$ and that are amenable to numerical integration of \eqref{II14} on a 
coarse mesh with grid-length $\Delta\sim \ell$ (e.g. see \cite{meneveau2000scale}). 

Onsager did not tackle this ``closure problem'' directly, but instead found a way to by-pass it. 
We discuss below his original approach using a ``point-splitting 
regularization'', but within our  ``block-spin'' regularization an analogous strategy may 
be followed. A key observation is that the stress-tensor $\btau_\ell(\bu,\bu)$ may be 
rewritten in terms of {\it velocity-increments} $\delta \bu(\br;\bx,t)=\bu(\bx+\br,t)-\bu(\bx,t),$ as 
\be \btau_\ell(\bu,\bu)=\langle \delta \bu\,\delta \bu\rangle_\ell
-\langle \delta \bu\rangle_\ell\,\langle \delta \bu\rangle_\ell, \lb{II19} \ee 
where $\langle f\rangle_\ell(\bx,t):=\int d^3r\, G_\ell(r) f(\br;\bx,t).$ This formula was 
originally obtained in Constantin et al. \cite{constantin1994onsager} in a slightly different form, 
and as above in \cite{eyink1995local} as a re-interpretation of their result. Equation \eqref{II19} 
is easy to verify by direct calculation, but it can be simply understood as the due to the 
invariance of the 2nd-order cumulant $\btau_\ell(\bu,\bu)$ to shifts of $\bu$ by 
vectors that are ``non-random'' with respect to the average $\langle\cdot\rangle_\ell$
over displacements $\br,$ i.e. that are independent of $\br.$ This allows $\bu(\bx+\br,t)$ 
in the definition \eqref{II13} of $\btau_\ell(\bu,\bu)$ to be replaced with $\delta\bu(\br;\bx,t),$
yielding the formula \eqref{II19}. Similarly, one may rewrite eq.\eqref{II7} for coarse-grained 
velocity-gradients in terms of increments as 
\be \grad_\bx\overline{\bu}_\ell(\bx,t)=-\frac{1}{\ell}\int d^3r\  (\grad G)_\ell(\br)\, \delta\bu(\br;\bx,t),  
\lb{II20} \ee 
using the fact that $\int d^3r\, (\grad G)_\ell(\br)=\bzed.$ The formulas \eqref{II19} and \eqref{II20},
together with the expression \eqref{II16} for local energy flux, are the main tools in the 
Onsager ``ideal turbulence'' theory for incompressible fluids. 

As an immediate application of these formulas, we can rederive the prediction of Onsager 
\cite{onsager1949statistical} 
that H\"older singularities $h\leq 1/3$ are required in the velocity field in order for energy dissipation 
to persist in the limit $\nu\to 0.$ Indeed, assuming for some constant $C>0$ that 
\be |\delta \bu(\br;\bx,t)|\leq C U (|\br|/L)^h, \lb{II21} \ee 
then it is straightforward to show using \eqref{II16},\eqref{II19} and \eqref{II20} that 
\be \Pi_\ell(\bx,t)=O(\ell^{3h-1}). \lb{II22} \ee
As is clear from \eqref{II15}, persistent energy decay at resolution 
length $\ell$ can only occur if $\int d^3x\, \Pi_\ell(\bx,t)\neq 0$ as $\nu\to 0.$ On the other hand, 
the resolution scale $\ell$ is completely arbitrary. For any fixed $\ell$ one can take $\nu$ sufficiently 
small so that the ``ideal equations'' \eqref{II12} or \eqref{II14} hold, and then subsequently 
further decrease $\ell.$  If the H\"older regularity \eqref{II21} held for all $(\bx,t)$ with $h>1/3,$
then clearly by \eqref{II22} it would follow that $\int d^3x\, \Pi_\ell(\bx,t)\to 0$ as $\ell\to 0.$
This is a contradiction, since the rate of decay of energy must be independent of the 
arbitrary length-scale of resolution $\ell$ as $\ell\to 0.$ Just as Onsager did, we thus 
infer that somewhere in the flow there must appear H\"older singularities $h\leq 1/3$ in the 
limit as $\nu\to 0$ or $Re\to \infty$.  

This prediction can be easily generalized within the Parisi-Frisch 
``multifractal model'' for the turbulent velocity field \cite{frisch1985singularity,uriel1995turbulence}
Using similar arguments as above, one can easily show that $p$th-order 
scaling exponents for ``velocity-structure functions'' 
\be S_p(\br)= \frac{1}{|V|}\int_V d^3x\, |\delta \bu(\br;\bx,t)|^p \sim C_p U^p (|\br|/L)^{\zeta_p}
\lb{II23} \ee 
must satisfy $\zeta_p\leq p/3$ for $p\geq 3.$ See \cite{constantin1994onsager,eyink1995local}
who took the limit $\nu\to 0$ first before then taking $\ell\to 0,$ and the more recent 
analysis of \cite{drivas2018onsagerincompressible} who take $\nu>0$ small but non-zero and 
exploit the arbitrariness of $\ell$ to derive $\zeta_p\leq p/3$ as a result on ``quasi-singularities'' 
of Navier-Stokes solutions.  
Recall within the multifractal framework that $h_p=d\zeta_p/dp$ gives the H\"older exponent 
that contributes dominantly to $\zeta_p=\inf_h \{ hp+ (3-D(h))\},$ with 
$D(h)$ the fractal dimension of the singularity set $S(h)$ on which H\"older exponent $h$ 
occurs. Because of the concavity of $\zeta_p$ in $p$ \cite{uriel1995turbulence,eyink2010notes},
one therefore concludes that $h_p\leq \zeta_p/p\leq 1/3$ for all $p\geq 3.$
Onsager's original result corresponds to the prediction that $h_{\min}=h_\infty\leq 1/3.$ 
These detailed predictions have been confirmed by laboratory experiments 
and numerical simulations. E.g. see \cite{uriel1995turbulence} for a survey 
or \cite{kestener2004generalizing} for more recent numerical results. 

It should be emphasized that the singularities inferred by this argument need not develop in finite time for 
Euler solutions starting from smooth initial data.  The most common experiments study turbulent flows 
produced downstream of wire-mesh grids in wind-tunnels or turbulent flows generated by flows past other 
solid obstacles, such as plates, cylinders, etc. \cite{sreenivasan1984scaling,pearson2002measurements}. 
The generation of turbulence is associated to vorticity fed into these flows by viscous boundary layers that 
detach from the walls. Since the boundary layers become thinner as $\nu = 1/Re$ decreases, the initial data 
of these experiments cannot be considered to be smooth uniformly in $\nu > 0$. Similar comments 
apply to numerical simulations. Long-time steady states with external body forcing correspond to taking 
first a limit $t\to\infty$ before subsequently taking $\nu\to 0.$ In that case, singularities have an infinite amount 
of time to reach the small ``dissipation scales" where viscosity is important. Only subsequently does one 
take $\nu = 1/Re\to 0$  so that the dissipation length shrinks to zero and the singularity becomes exact. 
In practice, some numerical simulations, such as that of \cite{kaneda2003energy}, show evidence 
that energy dissipation is anomalous when time-averaged over only a few large-eddy turnover times. However, 
a close examination reveals that those studies also do not employ initial data that is uniformly 
smooth as Reynolds number increases!  A standard practice is to initialize the simulation 
at high $Re$ by $\bu_\nu(\cdot,0) = \bu_{\nu'}(\cdot,T')$, where the second velocity field is the 
final state at time $T'$ of a smaller Reynolds-number $Re' < Re$ simulation performed at lower resolution 
and interpolated onto the finer grid of the $Re$-simulation (e.g. see p.L21 of \cite{kaneda2003energy}). 
This practice of ``nested'' initialization means that initial conditions $\bu_\nu(·, 0)$ have Kolmogorov-type spectra 
over increasing ranges of scales as $\nu$ decreases.

\section{Weak Euler Solutions and Dissipative Anomaly}\lb{sec:IId} 

A further observation of Onsager 
\cite{onsager1949statistical} was that any suitable (strong) limit $\bu_\star=\lim_{\nu\to 0} \bu$
of Navier-Stokes solutions with persistent energy dissipation as $\nu\to 0$ must 
correspond to a ``generalized'' Euler solution that dissipates kinetic energy. 
The notion of ``generalized'' solution proposed by Onsager corresponds exactly to the 
modern notion of a ``weak'' or ``distributional'' solution \cite{eyink1994energy,delellis2013continuous}. 
From our RG point of view, these are ``ultraviolet fixed-point solutions''
that are obtained by taking first the limit $\nu\to 0$ in the regularized equations \eqref{II9} 
to obtain (\ref{II12}) for $\bu=\bu_\star$ and then taking the UV limit $\ell\to 0$ so that
$\overline{\bu}_{\star\ell}\to \bu_\star,$ $\btau_\ell\to \bzed,$ and  
\be \partial_t \bu_\star + \grad_\bx\bdot (\bu_\star\bu_\star) = -\grad_\bx p_\star, \quad \grad_\bx\bdot \bu_\star=0 \lb{II24} \ee 
in the sense of distributions. 
Such ``weak'' or ``distributional'' Euler solutions possess the same self-similarity 
under rescalings $\bx'=\lambda \bx,$ $t'=\lambda^{1-h}t,$ $\bu'=\lambda^h\bu$
as do ordinary smooth Euler solutions \cite{uriel1995turbulence}. However, the kinetic energy balance 
for such dissipative weak solutions is modified by an ``anomaly term''. This result can be derived 
from the regularized energy balance (\ref{II15}) by taking the double limit first $\nu\to 0$ and 
then $\ell\to 0$ to obtain in the sense of distributions \cite{duchon2000inertial}
\bea 
&& \partial_t\left(\frac{1}{2}|\bu_\star|^2\right)+
\grad_\bx\bdot\left[\left( \frac{1}{2}|\bu_\star|^2+ p_\star\right)\bu_\star\right] 
=-\Pi_\star\cr
&&   \lb{II25} \eea
with 
\be \Pi_\star= -\lim_{\ell\to 0} \grad_\bx\overline{\bu}_{\star\ell}\,\bdots\,\btau_\ell(\bu_\star,\bu_\star).  \lb{II26} \ee 
The anomaly term is non-vanishing, $\Pi_\star\neq 0,$ when there is nonlinear energy flux 
$\Pi_\ell$ even as length-scale $\ell\to 0.$ 

As first noted by Polyakov \cite{polyakov1993theory,polyakov1992conformal} 
there is a striking analogy to conservation-law
anomalies in quantum field-theory, where terms similar to $\Pi_\star$ appear 
that vitiate conservation laws which hold classically. The most standard example 
is axial charge conservation which holds for a classical electrodynamic field coupled 
to a classical spinor field, but which is violated in quantum electrodynamics (QED). 
The source of that anomaly is a flux of axial/chiral charge produced at the 
ultraviolet cut-off momentum $\Lambda$ and which is transferred through momentum space 
to finite momentum values even as $\Lambda\to\infty$ (see Gribov \cite{gribov2001anomalies}). As remarked 
by Polyakov \cite{polyakov1992conformal} , ``in Kolmogorov's case the same happens with enstrophy or with energy.''
As \cite{polyakov1992conformal} also observed, the analogy of turbulent ``dissipative anomalies'' with the axial 
anomaly in QED is made more striking by the fact that Schwinger \cite{schwinger1951gauge} originally obtained 
the axial anomaly by a ``point-splitting regularization'' of UV divergences in QED, with a 
calculation formally very similar to that used by Kolmogorov \cite{kolmogorov1941dissipation}
in his derivation of the ``4/5th-law''  within his statistical theory of turbulence. Remarkably, we now know that 
Onsager in 1945 had performed a very similar ``point-splitting regularization'' of kinetic energy 
density 
\be \frac{1}{2}\bu(\bx,t)\bdot \overline{\bu}_\ell(\bx,t)
=\int d^3r\, G_\ell(\br) \bu(\bx,t)\bdot \bu(\bx+\br,t)  \lb{II27} \ee 
and took its time-derivative to derive a deterministic analogue of the ``4/5th-law'' 
(see \cite{eyink2006onsager} for a historical review). This calculation 
recovers the anomalous energy balance \eqref{II25} with an expression 
for the anomaly term that corresponds to the anisotropic 
version of the Kolmogorov ``4/5th-law'':  
\be \Pi_\star = \lim_{\ell\to 0} \frac{1}{4\ell}\int d^3r\, (\grad G)_\ell(\br)\bdot 
\delta \bu_\star(\br)|\delta \bu_\star(\br)|^2.   \lb{II28} \ee
See Duchon \& Robert \cite{duchon2000inertial} for a complete derivation
of \eqref{II25}, \eqref{II28} where it is also shown under reasonable assumptions that the anomaly 
term $\Pi_\star(\bx,t)$ coincides with the zero-viscosity limit $\varepsilon_\star(\bx,t)$ 
of the viscous energy dissipation \eqref{II4}. Note that there is no statistical 
averaging over ensembles of velocities in the formula \eqref{II28}, which gives a 
deterministic and space-time local version of the ``4/5th-law'' 
\cite{eyink2002local,taylor2003recovering}. 
This calculation was presumably the basis of the claims made about dissipative Euler solutions 
by Onsager \cite{onsager1949statistical}. 

It is still an open question in the mathematical foundations of Onsager's theory whether 
suitable limits  $\bu_\star=\lim_{\nu\to 0} \bu$ exist, which will yield the conjectured dissipative 
Euler solutions. Reasonable conditions which guarantee the existence of such Euler 
solutions as inviscid limits are verified over accessible ranges of Reynolds numbers 
\cite{constantin2017remarks,drivas2018onsagerincompressible,isett2017nonuniqueness}. 
Furthermore, in very deep mathematical 
work, dissipative, H\"older-continuous Euler solutions $\bu_\star$ have been constructed by ``convex 
integration'' methods, using ideas originating in the Nash-Kuiper theorem and Gromov's ``h-principle''  
\cite{delellis2012h,delellis2013continuous}. This circle of ideas led recently to a proof that Onsager's 
1/3 H\"older exponent is sharp and that dissipative Euler solutions exist with spatial H\"older 
exponent $(1/3)-\epsilon$ for any $\epsilon>0$ \cite{isett2016proof,buckmaster2017onsagers}. 
These dissipative Euler solutions $\bu_\star$ are not constructed by zero-viscosity limits but instead 
by an ``inverse RG'' procedure in which \eqref{II14} is solved for some specified $\overline{\bu}_{\ell_{k-1}}$
and $\btau_{\ell_{k-1}}$ and one then proceeds to a new length-scale $\ell_k \ll \ell_{k-1}$ by adding small-scale 
modes to the velocity field in such a way that $\overline{\bu}_{\ell_k}$ and $\btau_{\ell_k}$ 
again satisfy \eqref{II14} but with $|\btau_{\ell_k}|\ll |\btau_{\ell_{k-1}}|.$ Iterating this construction,
$\btau_{\ell_k}\to \bzed$ as $k\to\infty$ and the limit $\bu_\star=\lim_{k\to\infty}\bu_{\ell_k}$ is a weak Euler solution.  
 Further mathematical work along these lines will hopefully lead to more complete understanding of the 
 inviscid limit solutions  $\bu_\star=\lim_{\nu\to 0} \bu$ which can describe the infinite Reynolds-number 
 limit of physical turbulent flows, providing additional computational and theoretical tools.  

It should be strongly emphasized, however, that much of the Onsager theory does not depend 
upon the assumption that limits $\bu_\star=\lim_{\nu\to 0} \bu$ exist with viscosity taken to zero and 
the most significant empirical consequences follow whenever the Reynolds number is 
sufficiently large, but finite. This should be clear from the derivation of the bound $\zeta_p\leq p/3$
presented above, which never required the hypothesis that $\bu_\star=\lim_{\nu\to 0} \bu$ must exist 
(see also \cite{drivas2018onsagerincompressible}). The Onsager theory provides exact, 
non-perturbative tools for the analysis of fluid turbulence at very large (but finite) $Re.$ For example, 
the formulas \eqref{II19}, \eqref{II20} are the basis for a demonstration of the scale-locality of turbulent energy 
cascade at $Re\gg 1,$ whenever $0<\zeta_p<p$ for any $p\geq 3$ 
\cite{eyink2005locality}. As emphasized by Wilson (\cite{wilson1975renormalization},
Section VI), the property of locality by itself can provide very effective tools for systematic approximation.
In critical phenomena and quantum field-theory it was space-locality rather than scale-locality,
but the basic principle is the same. In \cite{eyink2006multi,eyink2006constitutive} 
the scale-locality of the turbulent stress 
$\btau_\ell(\bu,\bu)$ was exploited to develop a ``multi-scale gradient expansion''
which yields  systematic approximations to the turbulent stress tensor that can be the 
basis of practical closures and give physical insight. For example, these methods were 
applied to explicate the physical mechanism of inverse energy cascade in two-dimensional 
incompressible fluid turbulence \cite{chen2006physical,xiao2009physical}.  
Since the original work on energy cascade in incompressible neutral fluids, the Onsager theory 
has been applied to 2D enstrophy cascade \cite{eyink1995exact,eyink2001dissipation}, 
to 3D helicity cascade \cite{chae2003remarks,chen2003joint},  
to magnetohydrodynamic turbulence \cite{caflisch1997remarks,aluie2010scale,aluie2017coarse}
and to compressible fluid turbulence, both non-relativistic 
\cite{aluie2013scale,eyink2017cascades1} and relativistic \cite{eyink2017cascades2}. 
A new application is to entropy cascade and magnetic reconnection in kinetic turbulence 
of nearly collisionless plasmas \cite{eyink2018kinetic}. 

The Onsager theory predicts turbulent dissipative anomalies not only for Eulerian 
invariants such as energy, helicity, etc. but also in Lagrangian conservation laws
such as Kelvin's Theorem for circulations \cite{eyink2006Acirculations,eyink2006Bcirculations}
and Alfv\'en's Theorem for magnetic flux \cite{eyink2006breakdown}. A fundamental discovery 
in a 1998 paper of Bernard et al. \cite{bernard1998slow} is that Lagrangian particle trajectories 
are no longer unique for {\it exactly} specified initial data in the infinite-$Re$ limit, but instead
become ``spontaneously stochastic". As pointed out in  \cite{bernard1998slow} , this 
fascinating effect explains the enhanced mixing and dissipation of scalars (perfume, temperature 
fluctuations, etc.) advected by turbulent flows and is due to the breakdown of uniqueness 
of solutions to the initial-value problem for ODE's with only H\"older-continuous vector fields. 
This phenomenon thus depends essentially upon the H\"older singularities predicted by Onsager's 
analysis.  It has been shown recently that ``spontaneous stochasticity'' is the only possible 
mechanism for anomalous scalar dissipation away from solid walls  \cite{drivas2017lagrangian} 
and also provides a Lagrangian explanation for anomalous energy dissipation at Burgers
shocks  \cite{eyink2015spontaneous}. The deterministic conservation of Lagrangian invariants 
by smooth Euler solutions is thus no longer approached in the limit of $Re\to\infty$ but instead circulations
\cite{constantin2008stochastic,eyink2010stochastic} 
and magnetic fluxes \cite{eyink2007turbulent,eyink2009stochastic,eyink2013flux} 
are martingales (statistically conserved quantities) of the spontaneously stochastic flows. 

Onsager's pioneering ideas on ``ideal turbulence'' thus continue to stimulate new developments and provide, 
in the opinion of this author, the current ``standard model" of high-Reynolds turbulence. In a subject 
where nontrivial exact results are rare and where much work involves ad hoc closures and hand-waving 
phenomenology, it remains a central pillar of our understanding of fully-developed turbulent flows.

\bibliography{bibliography.bib}

\end{document}